\documentclass[
  aps,
  prx,
  reprint,
  superscriptaddress,
  amsmath,
  amssymb,
  longbibliography
]{revtex4-2}
\usepackage[normalem]{ulem}
\usepackage{graphicx}
\usepackage{dcolumn}
\usepackage{bm}
\usepackage{xcolor}
\usepackage{hyperref}
\usepackage{ulem}
\hypersetup{hidelinks}

\begin{document}

\title{Augmentation of cardiovascular flow measurements with generative diffusion models}
\author{Fabio Guglietta}\email{Contact author: f.guglietta@inail.it}
\affiliation{INAIL Research, Department of Occupational and Environmental Medicine, Epidemiology and Hygiene, Monte Porzio Catone, Italy.}
\author{Tianyi Li}
\affiliation{Tor Vergata University of Rome, Rome, Italy.}
\affiliation{INFN Tor Vergata, Rome, Italy.}
\author{Francesco Viola}
\affiliation{Gran Sasso Science Institute (GSSI), L'Aquila, Italy.}
\author{Luca Biferale}
\affiliation{Tor Vergata University of Rome, Rome, Italy.}
\affiliation{INFN Tor Vergata, Rome, Italy.}

\date{\today}

\begin{abstract}
{Cardiovascular imaging offers several ways to measure blood flow, but each provides an incomplete, resolution-limited, or noisy view of the underlying dynamics.} Reconstructing the missing information is inherently ambiguous because several flow fields may be compatible with the same observation. Here, we investigate generative diffusion models as tools for cardiovascular-flow generation and reconstruction using a fully controlled state-of-the-art dataset from a digital twin of a healthy human left heart. First, we show how unconditional samples reproduce the changing anatomy and the highly fluctuating statistical properties of the flow throughout the cardiac cycle. Second, we test the conditional inference of our generative model by reconstructing an entire 2D planar section of geometry and flow configurations from a limited portion. Repeated reconstructions 
{allow us to represent a range of plausible completions, going beyond a single deterministic estimate, while their mean retains the principal features of the reference flow}. 
{Comparison with a Gaussian process regression baseline on the same reduced domain highlights the main advantage of diffusion models: their reconstructed ensembles capture a broader range of velocity and gradient fluctuations, better encompassing ground-truth features that the baseline distributions can miss.}
Our results provide a quantitative proof of concept for probabilistic reconstruction of incomplete cardiovascular-flow measurements and demonstrate the potential for groundbreaking applications to the augmentation of \textit{in-vivo} cardiovascular imaging data.
\end{abstract}

\maketitle

\section{\label{sec:intro}Introduction}

{{Cardiovascular imaging reveals blood flow through measurements that are necessarily incomplete, owing to limited coverage, noise, and spatial or temporal averaging, which can leave important flow features unresolved.} Recovering the full flow -- at unseen spatial locations and unresolved spatial scales -- from partial data is an ill-posed problem: different velocity fields may agree with the available measurements. A probabilistic reconstruction can represent these alternatives and provide a basis for estimating instantaneous and statistical features that the instrument cannot resolve directly.}

{Limitations take different forms across cardiovascular imaging modalities. {Transthoracic echocardiography is portable and relatively inexpensive~\cite{malik2017_tte} and provides real-time images~\cite{stevens2024_dehazing}. By comparison, {time-resolved three-dimensional phase-contrast magnetic resonance imaging (4D-flow MRI)} acquires volumetric velocity data over several minutes, with a trade-off between acquisition time and spatial and temporal resolution~\cite{bissell2023_4dflow}. {Respiratory motion during acquisition can introduce additional artifacts in the measured velocity fields~\cite{bissell2023_4dflow}.}} {Color-Doppler echocardiography illustrates the incompleteness of flow measurements: it captures only the velocity component along the ultrasound beam~\cite{meyers2020_dover}. Aliasing and noise can corrupt the measured velocities~\cite{muth2011_dealiasing}, acoustic shadowing can obscure the color-Doppler signal~\cite{saric2026_artifacts}, and tissue clutter and the filtering used to suppress it can affect the available flow data~\cite{vixege2021_ivfm}.} More generally, missing or unreliable measurements can occur in regions where a complete velocity field is needed to characterize jets, vortices, and blood transport.}

Several reconstruction strategies have been developed to extract more complete hemodynamic information from these measurements. Signal-processing approaches can jointly denoise and dealias color-Doppler data~\cite{muth2011_dealiasing}, while intraventricular vector-flow methods (iVFM) infer unmeasured velocity components by imposing kinematic constraints{~\cite{vixege2021_ivfm}}. {The Doppler Velocity Reconstruction method uses a streamfunction--vorticity formulation to recover two-component ventricular flow fields. In tests using synthetic Doppler data derived from CFD, it was more robust to added noise and changes in probe placement than the two iVFM formulations compared in that study~\cite{meyers2020_dover}.} {More recently, AI-VFM has combined Doppler observations with planar mass and momentum conservation in a physics-informed neural network to reconstruct two-component velocity and spatial pressure fluctuations, including regions or time frames without color-Doppler input~\cite{maidu2025_aivfm}.} These developments demonstrate that neural networks and physical constraints can restore information absent from the measurement.  {These advances have been validated against simulated flow fields and illustrated with in vivo data~\cite{meyers2020_dover,vixege2021_ivfm,maidu2025_aivfm}, but recovering fine-scale fluctuations remains challenging. Vix\`ege et al.~\cite{vixege2021_ivfm} explicitly note that the spatial and temporal averaging inherent in color Doppler prevents their iVFM approach from resolving turbulent fluctuations. For AI-VFM, Maidu et al.~\cite{maidu2025_aivfm} report that spectral bias in the neural network smooths the reconstructed fields and leads to underestimation of peak pressure fluctuations. In the formulations presented in these studies, reconstruction yields a velocity-field estimate rather than a sampled conditional distribution of compatible fields~\cite{meyers2020_dover,vixege2021_ivfm,maidu2025_aivfm}. Here, {we take a step forward by investigating the problem with a probabilistic approach. Furthermore, we provide for the first time a systematic assessment of the entire range of turbulent velocity fluctuations, by examining } the resulting ensembles through velocity, gradient, and multiscale increment statistics.} Reconstructing a large missing region is intrinsically non-unique: several flow configurations may be compatible with the same limited observation. This ambiguity is {key in the presence of} turbulent flows, whose high-dimensional and chaotic dynamics allow multiple realizations within the missing region to be consistent with the available data~\cite{li2025stochastic}. 
{Patient-specific cardiac simulations reveal strongly intermittent velocity fluctuations over a broad range of time scales~\cite{guglietta2026lagrangian}, while conventional color-Doppler measurements average over spatial and temporal scales that can obscure fine-scale fluctuations~\cite{vixege2021_ivfm}.} A distribution of plausible fields could help estimate the likelihood of such unresolved events, which an average reconstruction alone cannot provide.

\begin{figure*}[th!]
    \centering
    \includegraphics[width=1.\linewidth]{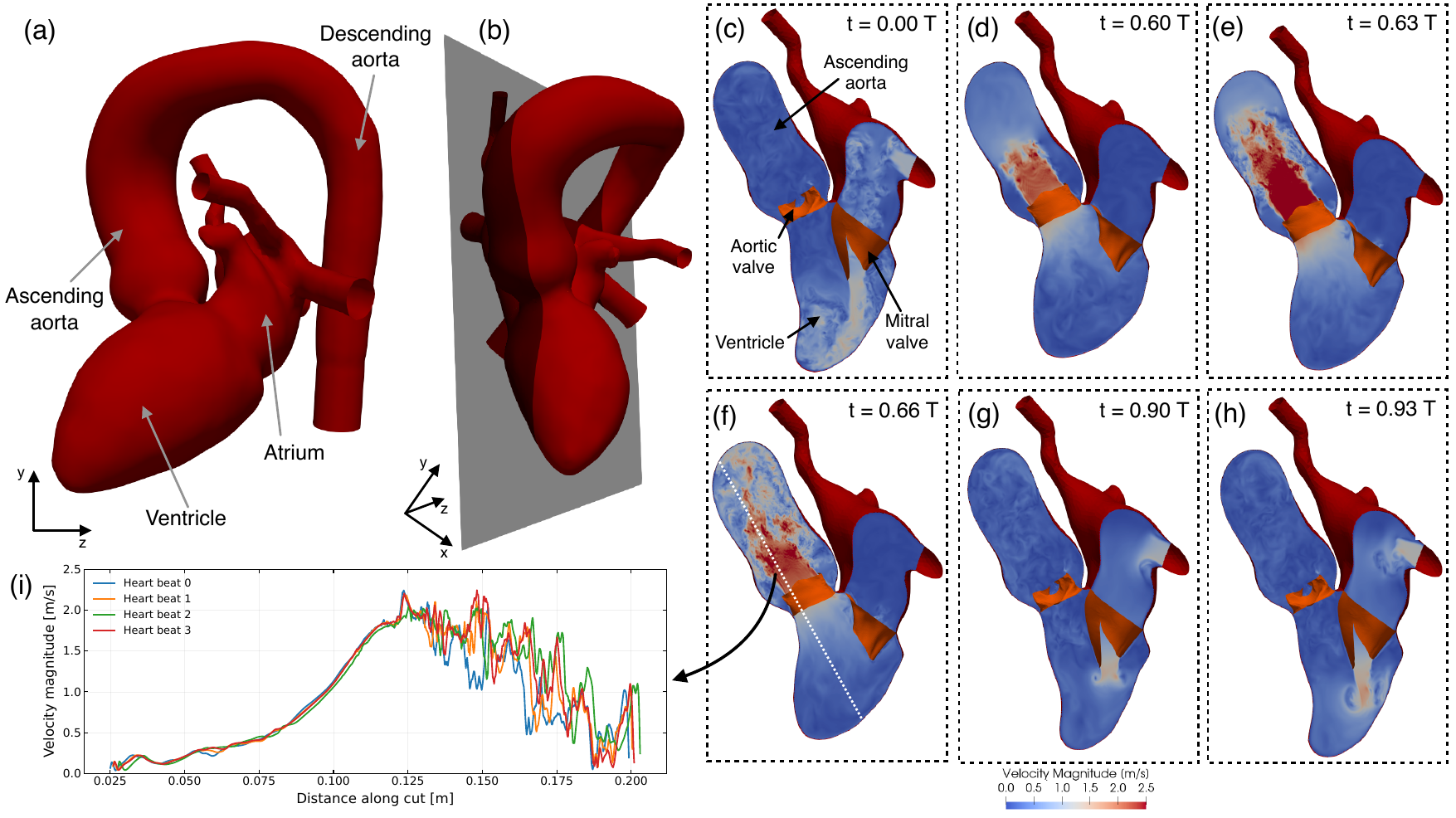}
    \caption{{Three-dimensional digital twin of the left heart and representative simulated flow. (a) Anatomical model showing the left atrium and ventricle and the ascending and descending aorta. (b) Section plane used to visualize the flow. (c--h) Flow fields at phases $t/T=0.00$, $0.60$, $0.63$, $0.66$, $0.90$, and $0.93$, respectively, where $T$ is the cardiac period and the phase origin is chosen at the configuration in (c). Colors indicate the velocity magnitude $v=|\mathbf{u}|$ in $\mathrm{m/s}$; the aortic and mitral valves are shown in orange. (i) Velocity magnitude along the dotted cut in (f) for four simulated heartbeats.}}
    \label{fig:sketch}
\end{figure*}

Generative models have meanwhile become increasingly prominent in cardiac-image synthesis and restoration. Generative adversarial networks have transformed anatomical models into realistic, labeled echocardiograms for training segmentation networks~\cite{gilbert2021_synthetic_echo}. Diffusion models conditioned on semantic maps have generated synthetic two-dimensional echocardiograms~\cite{stojanovski2023_echo_noise}. Cascaded video diffusion models have generated temporally coherent sequences conditioned on {a single reference frame and the left-ventricular ejection fraction}~\cite{reynaud2023_echo_diffusion}, while multimodally conditioned models have combined local structural cues with global text and image information to control cardiac-ultrasound video synthesis~\cite{zhou2024_heartbeat}. {Joint posterior sampling with separate diffusion priors for clean ultrasound and haze has also been used to reduce structured haze in cardiac ultrasound signals~\cite{stevens2024_dehazing}.}

{These studies synthesize or restore ultrasound images or signals and anatomical information, rather than reconstructing the underlying blood-velocity field. {In contrast,}  the flow-reconstruction task considered here targets the three velocity components together with the moving fluid domain. Its evaluation must therefore examine velocity fields and their spatial fluctuations, including chamber-scale circulation, localized jets, and turbulent features, rather than visual appearance alone.}

{Generative AI, particularly GANs and diffusion models, has been applied to turbulence for spatial and temporal reconstruction~\cite{buzzicotti2021_turbrot_inpainting,li2024_rotating_diffusion,li2025stochastic}, the generation of intermittent Lagrangian trajectories~\cite{li2024synthetic}, and physics-constrained synthesis of three-dimensional turbulent fields~\cite{li2026_physically_consistent_3d} (see Ref.~\cite{biferale2026generativeaiturbulencefoundations} for a recent review). Other contributions include GAN-based super-resolution reconstruction from unpaired turbulence data~\cite{kim2021_unsupervised_turbulence}, physics-informed diffusion for reconstruction from low-resolution or sparse observations~\cite{shu2023_pidm_flow}, and neural-field latent diffusion for conditional generation and data restoration in complex three-dimensional domains~\cite{du2024_confield}. Generative reduced-order modeling combines low-dimensional temporal evolution with diffusion-based recovery of full fields~\cite{gao2024_generative_forecasting}, while pretrained score-based models reconstruct and predict spatiotemporal dynamics from sparse, noisy measurements~\cite{li2024_pretrained_dynamics}. Recent work further assesses generative approaches to turbulent-flow super-resolution and sparse three-dimensional reconstruction~\cite{oommen2026_turbulentflows}. Together with advances in cardiac imaging, these studies motivate assessing diffusion models as probabilistic priors for incomplete cardiovascular-flow measurements, where moving anatomy and heterogeneous flow regimes pose additional challenges.}

Here, we adopt a two-step approach. First, we generate a large {\it ground-truth} database using a state-of-the-art digital twin in which both the flow and the missing information are fully controlled. Second, we extract aortic and ventricular slices from this database and use different heartbeats for training and testing.
{Our digital twin builds on a fluid--structure--electrophysiology interaction framework that couples cardiac electrical activation, tissue mechanics, and blood flow~\cite{viola2023high}. The left-heart anatomy and simulation setup used here follow the patient-specific model described in Ref.~\cite{guglietta2026lagrangian}, providing high-resolution three-component velocity fields and the evolving fluid domain throughout the heartbeat.} {The simulations couple the electrical activation of the myocardium to contraction of the cardiac walls, while the moving mitral and aortic valves interact with the blood flow in both directions: their motion shapes the flow, and fluid forces act back on the valve leaflets. The resulting changes in chamber and vessel geometry, together with valve opening and closure, produce strongly time-dependent flow patterns across the cardiac cycle, as illustrated in Fig.~\ref{fig:sketch}. {{Panels (c--h) show the spatial distribution of velocity magnitude at representative phases of the cardiac cycle.} The profiles in (i), evaluated along the cut in (f) for four heartbeats, retain a similar large-scale shape while differing in their local fluctuations.}} {This study builds on a unique simulation database combining fully resolved three-dimensional velocity and pressure fields with the evolving geometry of the heart and valves over 50 heartbeats. This combination captures both the coupled flow–anatomy dynamics within each heartbeat and the flow variability across successive cycles. The simulation setup and the extraction of the two-dimensional datasets used here are described in Sec.~\ref{sec:methods}.} 

{In this paper we present two main results: (i) unconditional generation of aortic flow slices reproduces the flow and geometry variability (Fig.~\ref{fig:unconditional_snapshots}); (ii) conditional sampling reconstructs missing regions in aortic and ventricular slices from a known strip containing the three velocity components (Fig.~\ref{fig:sketch_recon_aorta}). Comparison with a Gaussian process regression baseline evaluates reconstruction profiles and ensemble statistics in a reduced aortic domain (Fig.~\ref{fig:gpr_stripe}).}

{Our study provides a simulation-based proof of concept for probabilistic augmentation of incomplete cardiovascular flow fields; application to in-vivo measurements remains a future validation step.}

\section{Methods}
\label{sec:methods}

\subsection{Fluid--structure--electrophysiology interaction (FSEI) simulations}
\label{sec:fsei}

The flow database is obtained from high-fidelity simulations of a patient-specific left-heart model using the fully coupled fluid--structure--electrophysiology interaction (FSEI) framework described and validated in Refs.~\cite{viola2020fsei,viola2022fsei_gpu,viola2023fsei,guglietta2026lagrangian}. The anatomical model was reconstructed by semiautomatic segmentation in 3D Slicer~\cite{3dslicer} from a thoracic computed-tomography scan of a healthy 44-year-old male, acquired during diastole with a spatial resolution of $0.728\times0.728\times1.0\,\mathrm{mm}^3$. It comprises the left atrium, left ventricle, mitral and aortic valves, ascending aorta, aortic arch, and the initial portion of the descending aorta. The database considered here corresponds to the healthy physiological configuration at a heart rate of $\text{HR}=60\,\mathrm{bpm}$. The pulmonary veins provide the inflow boundaries, while the aortic outlet is coupled to a Windkessel model that establishes the physiological arterial-pressure range. Further details of the anatomical reconstruction and physiological setup are given in Refs.~\cite{viola2023high,guglietta2026lagrangian}.

Blood is modeled as an incompressible Newtonian fluid of density $\rho=1000\,\mathrm{kg\,m^{-3}}$ and kinematic viscosity $\nu=4.8\times10^{-6}\,\mathrm{m^2\,s^{-1}}$. Its velocity $\mathbf{u}$ and pressure $p$ satisfy
\begin{align}
&\partial_t\mathbf{u}+(\mathbf{u}\!\cdot\!\boldsymbol{\nabla})\mathbf{u}
=-\frac{\boldsymbol{\nabla}p}{\rho}+\nu\nabla^2\mathbf{u}+\mathbf{f}\ ,
\label{eq:navier_stokes}\\
&\boldsymbol{\nabla}\!\cdot\!\mathbf{u}=0\ ,
\label{eq:continuity}
\end{align}
where $\mathbf{f}$ is the immersed-boundary forcing that imposes no slip on the moving cardiac surfaces. The equations are discretized with second-order finite differences on a staggered Cartesian grid containing $630\times720\times630$ points over a $0.21\times0.24\times0.21\,\mathrm{m}^3$ domain. The resulting uniform spatial resolution is $\Delta x=3.33\times10^{-4}\,\mathrm{m}$ and the time step is $\Delta t=3\times10^{-6}\,\mathrm{s}$. Fluid--surface coupling is enforced through a moving-least-squares immersed-boundary method~\cite{deTullio2016MLS}.

The cardiac walls and valve leaflets are represented by deformable triangulated meshes whose in-plane elasticity and bending resistance are described by a spring-network model~\cite{viola2020fsei,viola2022fsei_gpu}. {The mitral and aortic valves are highlighted in Fig.~\ref{fig:sketch}(c--h).} {Resolving their motion is essential for the velocity field: wall deformation changes the shape and volume of the flow domain, while valve opening and closure control the flow pathways and the formation of jets.} Myocardial activation is computed with a bidomain electrophysiology formulation and converted into active tension~\cite{delcorso_2022_fast_ef,viola2023high}. At every time step, the structural nodes are driven by the combined active, elastic, and hydrodynamic forces, while their motion updates the immersed boundaries seen by the fluid. This two-way coupling allows the simulations to reproduce the cardiac-cycle-dependent motion of the chambers and valves and, consequently, the associated changes in both flow and geometry {(see Fig.~\ref{fig:sketch})}. The simulations are GPU accelerated, and three cardiac cycles are discarded before collecting data to remove the initial transient and reach a periodic regime~\cite{viola2022fsei_gpu,guglietta2026lagrangian,vagnoli2026ibflowsopensourcemultigpuimmersed}.

\subsection{Database construction}
\label{sec:fsei_database}

{The FSEI database contains the full three-dimensional velocity and pressure fields and the evolving meshes of the heart and valves. To construct two-dimensional samples, we extract fields on four parallel $yz$ planes and select separate aortic and ventricular crops from each plane. Each sample thus represents one anatomical region rather than the entire left heart. The four planes have distinct $x$ coordinates that remain fixed for all sampling instants and heartbeats. They are selected around a section intersecting the valves to include regions with complex flow structures, with adjacent planes separated by approximately five Eulerian grid spacings. Fields are extracted at 100 instants per retained heartbeat. We focus on the aortic and ventricular regions to assess flow generation and reconstruction in two anatomically distinct domains with different flow characteristics. The crops contain $176\times416$ and $304\times272$ grid points, respectively, corresponding to physical dimensions of approximately $59\times139\,\mathrm{mm}^2$ and $101\times91\,\mathrm{mm}^2$.}

{Each sample has four input channels: the three Cartesian velocity components, $u_x$, $u_y$, and $u_z$, and a binary geometry indicator $\chi$. Before normalization, $\chi=1$ at fluid points and $\chi=0$ outside the fluid domain. This fourth channel encodes the instantaneous shape of the fluid region within the slice.} The velocity magnitude reported in the figures is computed as 
\begin{equation}\label{eq:vel_mag}
    v =\sqrt{u_x^2+u_y^2+u_z^2}\ .
\end{equation}
{Pressure and the structural meshes are retained in the full database but are not input channels for the diffusion models studied here.} Before being supplied to the diffusion model, {the four-channel samples $(u_x,u_y,u_z,\chi)$ are rescaled} to the interval $[-1,1]$ { using min-max normalization: a value $x$ is mapped to $2(x-x_{\min})/(x_{\max}-x_{\min})-1$, where $x_{\min}$ and $x_{\max}$ are the minimum and maximum values used for that channel}; generated and reconstructed fields are mapped back to their original physical ranges for the analyses reported below.

{The database spans 50 retained heartbeats, partitioned into 25 for training and 25 for testing. Each heartbeat contributes $4\times100$ samples per anatomical region, yielding $10^4$ training samples and $10^4$ test samples for each region. Splitting the data by heartbeat ensures that no test field is presented to the model during training. Preliminary tests varying the number of training images indicated that model performance had saturated at $10^4$ training samples (data not shown); this value was therefore adopted as the training-set size. The choice of the number of optimization iterations is assessed in Appendix~\ref{sec:diffusion_method}. We use ground truth (GT) to denote the simulation fields, specifying ``training GT'' or ``test GT'' when the distinction matters. The reference statistics used to validate unconditional generation are evaluated on all $10^4$ aortic test GT fields.}

\begin{figure*}[t!]
    \centering
    \includegraphics[width=1.\linewidth]{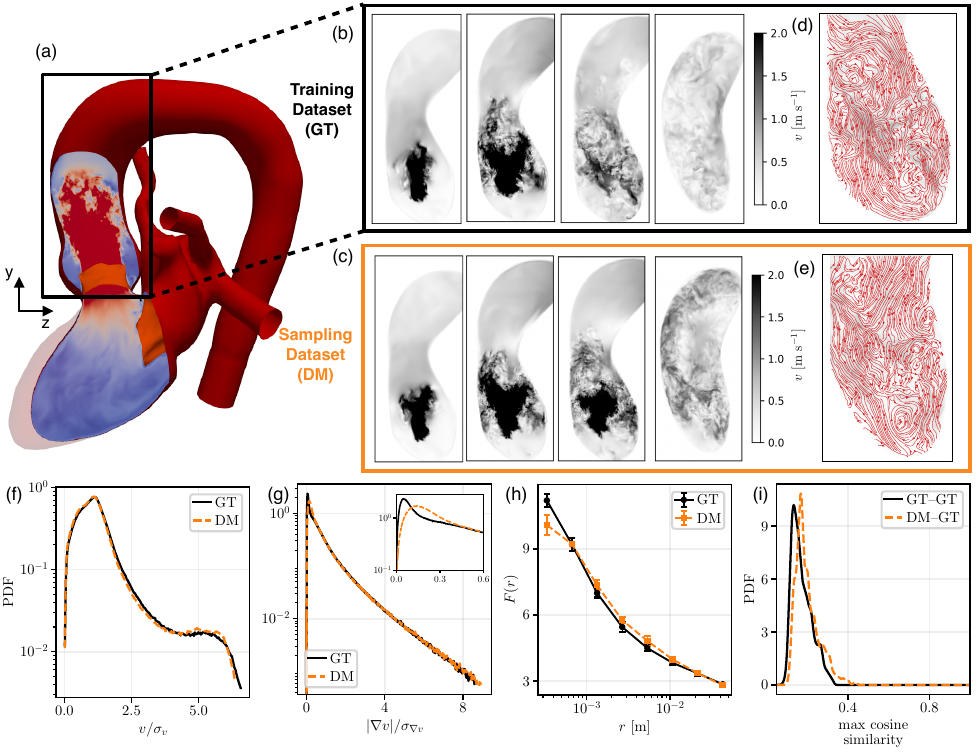}
    \caption{{Aortic slices and unconditional generation. (a) Cardiac geometry with the selected aortic region outlined in black. {(b) Four ground-truth (GT) slices selected at random from the training dataset. (c) Four fields selected independently at random from the unconditionally generated diffusion-model (DM) ensemble. The GT and DM snapshots are unpaired: their positions in the panels do not indicate correspondence.} Grayscale indicates the velocity magnitude $v$ in $\mathrm{m\,s^{-1}}$. {(d,e) Red streamlines of the velocity projected onto the slice plane for GT and DM snapshots, respectively, with a zoom near the aortic valve.} (f-g) PDFs of $v/\sigma_v$ and $|\nabla v|/\sigma_{\nabla v}$, respectively, where the gradient is evaluated in the slice plane; the inset enlarges the PDF peak. Each PDF is rescaled using the standard deviation of the corresponding quantity in its own ensemble. (h) Flatness $F(r)$ of vertical speed increments, as defined in Eq.~\eqref{eq:flatness}, with $r$ in meters. {Symbols show the midpoint of the mean flatness values obtained from two disjoint random halves of each ensemble; error bars span the interval between these two values.} (i) PDF of the maximum cosine similarity (defined in Eq.~\eqref{eq:cosinesim}) between velocity-magnitude gradient fields: generated fields versus training GT fields (DM--GT, orange dashed) and distinct training GT fields versus one another (GT--GT, black solid).}}
    \label{fig:unconditional_snapshots}
\end{figure*}

\section{Results}
\label{sec:results}

\subsection{Unconditional generation}
\label{sec:unconditional_generation}
{A diffusion model (DM) learns to generate new fields by reversing a gradual corruption process~\cite{ho2020denoising}. During training, noise is added to complete examples and a neural network learns to remove it at different noise levels. Once trained, the model starts from random noise and progressively transforms it into a new field consistent with the patterns learned from the data. Different noise realizations yield different samples, allowing the model to represent a distribution of possible fields rather than a single prediction. The architecture, training objective, and sampling procedure are detailed in Appendix~\ref{sec:diffusion_method}.}

{We first test whether this approach reproduces the variability of complete aortic flow fields. The model is trained on the $10^4$ aortic slices described in Sec.~\ref{sec:fsei_database} and jointly generates the three velocity components and the fluid-domain geometry. Generation is unconditional: no reference field, partial observation, or cardiac phase is supplied. The sampled ensemble must therefore reproduce both the spatial organization of the flow and its variability across the cardiac cycle.}

Figure~\ref{fig:unconditional_snapshots}(a) illustrates the cardiac geometry and the location of {an aortic slice extracted from the full three-dimensional database} (the selected region is identified by the {black rectangle}). {Figure~\ref{fig:unconditional_snapshots}(b,c) compares four ground-truth (GT) aortic slices selected at random from the training dataset with four fields selected independently at random from the unconditionally generated DM ensemble. Each generated field originates from an independent noise realization, without conditioning on any GT snapshot. The images in the two panels are therefore unpaired and are not matched by cardiac phase or spatial configuration.} The selected fields span the range from relatively quiescent configurations to states characterized by intense, localized flow, consistent with different stages of the cardiac cycle. In addition to the range of velocity magnitudes and the spatial localization of the systolic outflow, the generated samples reproduce the variations in aortic geometry that accompany the cardiac cycle. In particular, the aortic expansion associated with systolic ejection produces appreciable changes in the shape and extent of the fluid domain, which are also evident in the sampled fields. {The DM thus generates plausible velocity patterns together with the associated changes in aortic shape, capturing the joint variability of flow and geometry across the cardiac cycle.}

{Figure~\ref{fig:unconditional_snapshots}(d,e) shows streamlines of the projected velocity field for GT and generated snapshots, respectively, with a zoom near the aortic valve to highlight local recirculation and changes in flow direction. These are instantaneous streamlines of the in-plane velocity.}

{To evaluate the generated ensemble beyond visual plausibility, we examine three complementary properties: the distributions of speeds $v$ (see Eq.~\eqref{eq:vel_mag}) and spatial speed gradients, the scale dependence of speed fluctuations, and the diversity of spatial flow structures.}

{
The probability density function (PDF) of $v$ describes how often different speeds occur across fluid-domain points in the ensemble, without retaining their spatial locations. Figure~\ref{fig:unconditional_snapshots}{(f)} shows the PDF of $v/\sigma_v$, where $\sigma_v$ is the standard deviation of speed in the corresponding ensemble, for both DM and GT. 
The two PDFs closely agree in shape, including the peak and the decay toward high speeds.}

{To probe local spatial variations, we next consider $\nabla v(\mathbf{x})$, where $\nabla$ acts within the slice plane. Gradients are evaluated in the interior of the fluid domain, excluding stencils that cross its boundary. Figure~\ref{fig:unconditional_snapshots}{(g)} compares the PDFs of $|\nabla v|/\sigma_{\nabla v}$, with $\sigma_{\nabla v}$ the standard deviation in each ensemble. The distributions closely follow one another over much of their range, including the large-gradient tail. The inset resolves differences near the peak that are less apparent in the full-range plot.}
{To characterize multiscale spatial fluctuations, we use increments of $v$ along vertical lines in the slice. Denoting the unit vector in this direction by $\mathbf{e}_{\mathrm{vert}}$, we define}
\begin{equation}
{\delta_r v(\mathbf{x}) = v(\mathbf{x}+r\mathbf{e}_{\mathrm{vert}})-v(\mathbf{x}),}
\end{equation}
and their scale-dependent flatness,
\begin{equation}
{F(r)=\frac{\left\langle\left(\delta_r v\right)^4\right\rangle}
{\left\langle\left(\delta_r v\right)^2\right\rangle^2}.}
\label{eq:flatness}
\end{equation}
{Here, the averages are taken over all pairs of points within the fluid domain of each slice separated by $r$ in the vertical direction.} {{The increment $\delta_r v$ measures the change in speed between these points,} while the flatness compares the frequency of unusually large changes with that expected from the typical fluctuations at the same separation: because the numerator contains the fourth power, rare sharp changes contribute strongly. Tracking $F(r)$ therefore tests whether the model reproduces intermittency across spatial scales, not merely the overall distribution of speeds.}
{In Fig.~\ref{fig:unconditional_snapshots}{(h)}, the separation $r$ is plotted in meters; the smallest separation is approximately $3.33\times10^{-4}\,\mathrm{m}$.}
{Flatness is calculated separately for each image. Each ensemble is randomly divided, without replacement, into two equally sized subsets, and the mean flatness is evaluated within each subset at every separation $r$. Symbols show the arithmetic mean of these two values, and error bars span the interval between them.}
{As shown in Fig.~\ref{fig:unconditional_snapshots}{(h)}, both ensembles exhibit larger flatness at small separations, reflecting the strong contribution of intense local speed increments to the fourth-order moment. The flatness decreases with increasing separation.} {Generated and GT curves exhibit a similar scale dependence, with closer agreement at larger separations.} The model thus reproduces not only the distribution of $v$, but also the scale dependence of its intermittent spatial fluctuations.

\begin{figure*}[t!]
    \centering
    \includegraphics[width=1\linewidth]{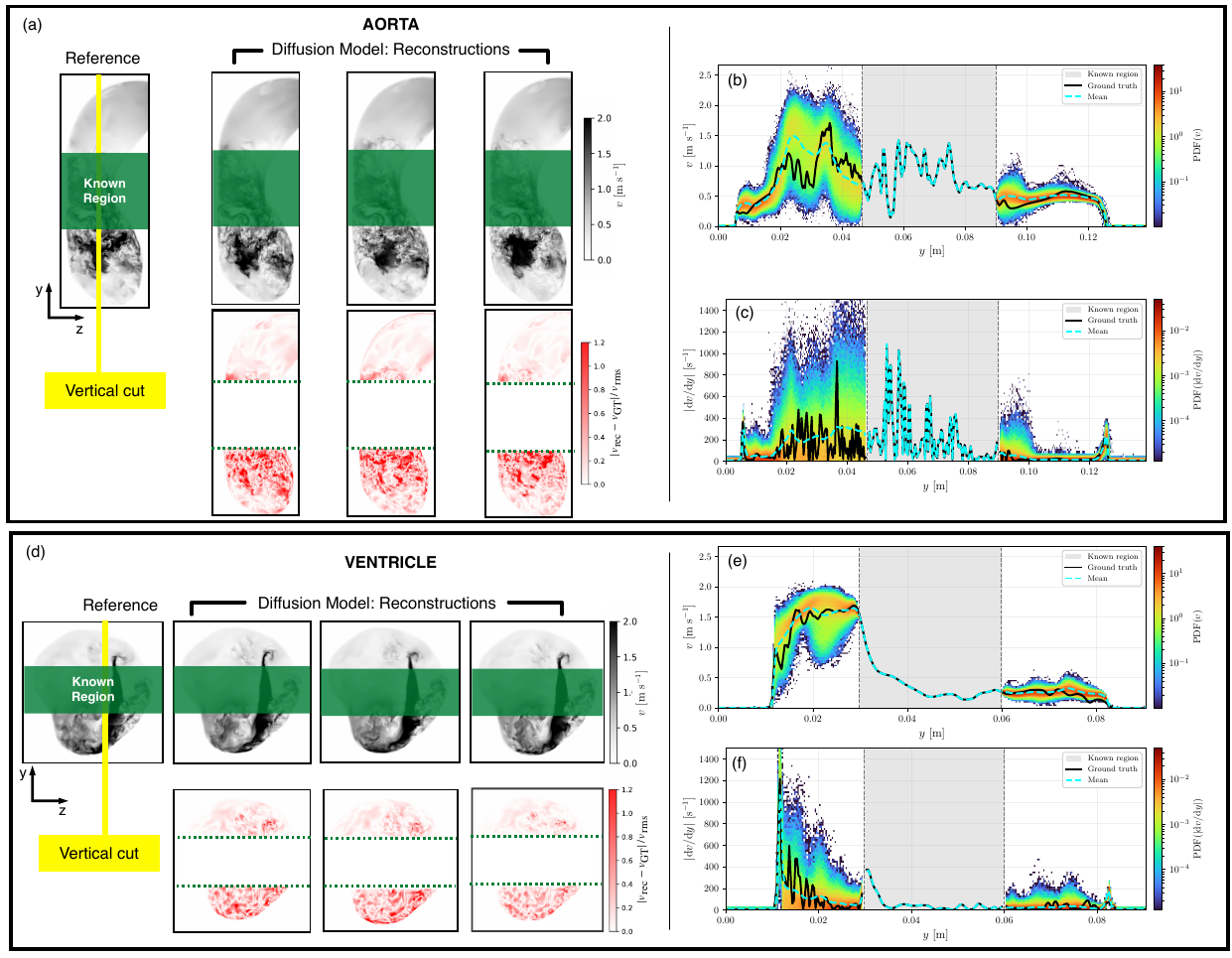}
    \caption{{Conditional reconstruction of aortic (upper group) and ventricular (lower group) flow fields. (a,d) A GT slice and three stochastic DM reconstructions of the same slice, with the supplied central strip shaded in green. Grayscale indicates the velocity magnitude $v$ in $\mathrm{m\,s^{-1}}$. The red maps show the absolute speed error normalized by the spatial RMS speed of the corresponding GT slice, $e_k=|v_{\mathrm{rec}}^{(k)}-v_{\mathrm{GT}}|/v_{\mathrm{rms}}$, where $v_{\mathrm{rms}}=\sqrt{\langle v_{\mathrm{GT}}^2\rangle_\Omega}$ and $\Omega$ is the full GT fluid domain. Yellow lines mark the vertical cuts used for the profiles. (b,e) Speed along the aortic and ventricular cuts, respectively. (c,f) Absolute directional speed gradient $|\mathrm{d}v/\mathrm{d}y|$ along the same cuts. Black curves show GT and dashed cyan curves show the ensemble mean. Colors represent pointwise PDFs across the stochastic reconstructions on a logarithmic density scale; gray bands mark the known intervals. Gradients are computed separately for each reconstruction before ensemble averaging, excluding finite-difference stencils that cross known--missing interfaces. {The horizontal axis reports the vertical coordinate $y$ along the cut, in meters.}}}
    \label{fig:sketch_recon_aorta}
\end{figure*}

{The close agreement between the generated and GT statistics in Fig.~\ref{fig:unconditional_snapshots}(f--h) makes it important to distinguish generalization from memorization. Here, generalization refers to generating new flow configurations that reproduce the statistical properties of the reference ensemble, rather than memorizing individual training fields; it does not imply transfer to different anatomies or physiological conditions. Agreement in PDFs and flatness alone cannot distinguish these possibilities, motivating a comparison of the spatial structures of generated and training fields.} {We use cosine similarity between velocity-magnitude gradient fields as a diagnostic of structural similarity, rather than a general measure of generative quality.} We therefore compare the spatial organization of the velocity-magnitude gradients in each generated field with that of the {training GT} fields. Using gradients rather than the velocity magnitude itself reduces the influence of the common geometry and large smooth regions, making the comparison more sensitive to the position and shape of jets, and smaller-scale flow structures. For a query field $q$, its maximum cosine similarity with the {training GT set} is defined as
\begin{equation}\label{eq:cosinesim}
C_{\max}(q)=\max_{j\in{\mathrm{GT}_{\mathrm{train}}}}
\frac{
\left\langle\nabla v_q,\nabla v_j\right\rangle
}{
\left\|\nabla v_q\right\|_2
\left\|\nabla v_j\right\|_2
}.
\end{equation}
{For two gradient fields $\mathbf{a}$ and $\mathbf{b}$, the inner product is $\langle\mathbf{a},\mathbf{b}\rangle=\sum_{\mathbf{x}\in\mathcal{G}}\mathbf{a}(\mathbf{x})\cdot\mathbf{b}(\mathbf{x})$, where $\mathcal{G}$ contains all grid points in the slice and $\cdot$ is the dot product of the in-plane gradient vectors; the corresponding norm is $\|\mathbf{a}\|_2=\sqrt{\langle\mathbf{a},\mathbf{a}\rangle}$.}
{For each generated field, we calculate its similarity to every training GT field and retain the largest value: this identifies the closest training example and tests whether the model may be reproducing fields it has seen.}
A value close to one means that two fields have almost the same spatial arrangement of velocity gradients, whereas a lower value indicates more distinct flow structures. As a reference, the same comparison is performed {between training GT fields}, excluding the query field itself from the maximization. A concentration of {DM--GT} similarities near one, or values systematically larger than the {GT--GT} similarities, would indicate that the generated ensemble contains less structural diversity than {distinct training fields}.

{Figure~\ref{fig:unconditional_snapshots}{(i)}} instead shows that the two distributions have comparable support and similar shapes, with only a small shift of the {DM--GT} distribution toward larger values and no accumulation near one. The generated fields are therefore {not systematically closer to individual training examples than distinct training fields are to one another}. Together, the one-point PDF, flatness, and cosine-similarity analyses show that the unconditional model captures the statistical variability, multiscale spatial organization, and structural diversity of the aortic flow database.
\subsection{Conditional generation}
\label{sec:conditional_generation}
We next move to our second result and assess the diffusion model as a conditional generative prior for reconstructing incomplete cardiovascular flow fields. The {full-slice} conditional reconstruction uses the same neural network and the same trained parameters as the unconditional generation; no additional training is performed. {The aortic and ventricular fields considered here are selected from their respective test sets and were not used for training; the dimensions of the complete slices are given in Sec.~\ref{sec:fsei_database}.} In the reconstruction configuration considered in this section, only the central horizontal region shaded in green in Fig.~\ref{fig:sketch_recon_aorta} is supplied to the model. {This strip spans the full horizontal extent and contains $176\times128$ grid points for the aorta and $304\times88$ for the ventricle, with physical dimensions of approximately $59\times43\,\mathrm{mm}^2$ and $101\times29\,\mathrm{mm}^2$, respectively. It covers approximately $30\%$ of each domain.} The present configuration is deliberately stringent because it removes two blocks that together account for approximately $70\%$ of the complete field. In practical applications, unavailable measurements would more commonly be distributed over several smaller, disconnected portions of the domain, although their combined area could still approach $70\%$ of the full field. {For each selected field, the two missing regions are reconstructed $4\times10^3$ times from independent initial noise fields, always conditioning on the same observed strip.}
{Starting from random noise, the model progressively removes noise to generate a complete flow field. During this iterative process, diffusion posterior sampling (DPS)~\cite{chung2023diffusion} adjusts the evolving reconstruction to reduce the mismatch between the model's estimate and the supplied data within the known strip. The technical details are given in Appendix~\ref{sec:diffusion_method}.}

\begin{figure*}[t!]
    \centering
    \includegraphics[width=1.\linewidth]{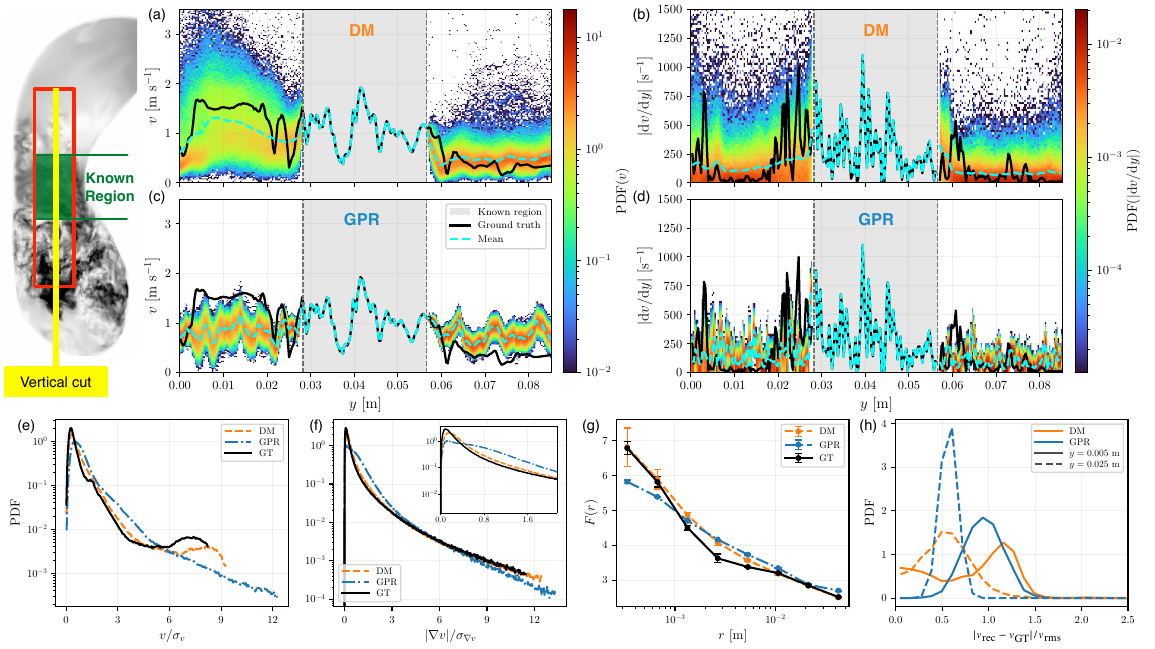}
    \caption{{Comparison of diffusion model (DM) and Gaussian process regression (GPR) reconstructions in the reduced aortic domain of $32\times256$ points, outlined in red on the reference slice. Green lines bound the known strip and the yellow line marks the vertical cut. (a,c) Three-component speed $v$ along the cut for DM and GPR, respectively. (b,d) Absolute directional speed gradient $|\mathrm{d}v/\mathrm{d}y|$ along the same cut. Black curves show GT, dashed cyan curves show ensemble means, and colors encode pointwise probability densities on a logarithmic scale. Gray shading marks the known interval; derivative stencils crossing its boundaries are excluded. (e,f) Ensemble PDFs of $v/\sigma_v$ and $|\nabla v|/\sigma_{\nabla v}$, respectively, evaluated over the fluid domain of the full crop for different GT images and their corresponding DM and GPR reconstructions. In (f), $\nabla v$ is the in-plane gradient and the inset enlarges the PDF peak. Each PDF uses the standard deviation of its own ensemble. (g) Flatness of vertical speed increments (see Eq.~\eqref{eq:flatness}). In (e-g), DM is orange dashed, GPR blue dash-dotted, and GT black solid. (h) PDFs of the absolute speed error $e=|v_{\mathrm{rec}}-v_{\mathrm{GT}}|/v_{\mathrm{rms,GT}}$ at $y=0.005\,\mathrm{m}$ (solid) and $y=0.025\,\mathrm{m}$ (dashed), across reconstructions of the fixed reference slice in (a-d). Both methods and positions use the same GT RMS speed, evaluated over all grid points of the full crop, including the known region. DM is orange and GPR blue; each PDF has unit area.}}
    \label{fig:gpr_stripe}
\end{figure*}

Figure~\ref{fig:sketch_recon_aorta} provides a qualitative assessment of the conditional reconstructions for one aortic and one ventricular {GT} slice. {The aortic field is selected near peak systole, whereas the ventricular field corresponds to early diastole, when the mitral valve opens and ventricular filling begins. These phases are selected to assess reconstruction in the presence of strong ejection and filling jets, respectively.} {The velocity field within the green strip is supplied as the observation, while the regions above and below the strip are generated by the conditioned model.} Despite the limited observation, the reconstructions recover the overall geometry of each fluid domain and the dominant organization of the corresponding velocity field.

{For each reconstructed field $k$, the dimensionless error map is defined as}
\begin{equation}
{e_k(\mathbf{x})=
\frac{\left|v_{\mathrm{rec}}^{(k)}(\mathbf{x})-v_{\mathrm{GT}}(\mathbf{x})\right|}
{v_{\mathrm{rms}}}\ ,}
\label{eq:reconstruction_error}
\end{equation}
{with $v_{\mathrm{rms}}=\sqrt{\left\langle v_{\mathrm{GT}}^2\right\rangle_{\Omega}}$}.
{Here, $\langle\cdot\rangle_{\Omega}$ denotes a spatial average over all fluid grid points of the GT slice, including both the known and missing regions. The same scalar normalization is used at every point and for every reconstruction of that slice.}

For the aorta, the DM preserves the elongated shape of the domain and reconstructs the marked contrast between the relatively smooth upper region and the highly heterogeneous lower region. The latter contains the aortic valve and the systolic jet, whose sharp velocity gradients and multiscale fluctuations make this portion of the field particularly challenging. The stochastic samples reproduce the location and spatial extent of this intense flow region, while differing in its finer structures. Consistently, {the normalized error maps defined in Eq.~\eqref{eq:reconstruction_error}} show their largest values around the valve, the jet, and the associated localized gradients, whereas the smoother upper region is reconstructed more uniformly.

The ventricular slice is similarly challenging. The reconstructions preserve the geometry and recover the dominant high-velocity structure on the right-hand side of the field. Differences among the samples are again concentrated around the more finely structured flow features and near sharp spatial variations, while the large-scale organization remains consistent with the reference. Figure~\ref{fig:sketch_recon_aorta} therefore shows that the known strip strongly constrains the global configuration of the missing field, but does not uniquely determine all of its small-scale features.

{This remaining variability is a consequence of incomplete observation, not a defect of the reconstruction: multiple turbulent-flow configurations can be compatible with the same measured strip. A deterministic estimate can summarize the missing field but cannot represent this range of possible completions. Guided conditional sampling instead provides an ensemble {that represents variability and the occurrence of intense flow within the sampled completions.} If validated against clinical measurements, such probabilistic information could eventually help identify flow-based markers of cardiovascular risk.}
To quantify how these reconstructions vary around the reference field, we next compare their velocity magnitudes along one-dimensional cuts through the reconstructed domains.

{Figure~\ref{fig:sketch_recon_aorta}(b,e) shows the speed profiles along the yellow vertical cuts through the aortic and ventricular slices, respectively. The gray bands identify the known intervals, where the reconstructed profiles coincide with the GT by construction. Outside these intervals, the color maps represent the pointwise distributions of reconstructed speeds, with black curves indicating the GT and cyan curves the ensemble mean.}

{At each coordinate $y$, the speeds $v_{\mathrm{rec}}^{(k)}(y)$ from the independent stochastic reconstructions are binned to estimate their PDF. Colors encode the probability density on a logarithmic scale: warm colors indicate more concentrated values, while blue and purple identify sparsely populated parts of the distribution.}

{The corresponding gradient profiles in Fig.~\ref{fig:sketch_recon_aorta}(c,f) show $|\mathrm{d}v/\mathrm{d}y|$, computed separately for every reconstruction. Their ensemble mean is therefore the mean absolute gradient, rather than the absolute gradient of the mean speed. The broad distributions on the smaller-$y$ side of the known intervals show that reconstructions agreeing on the main speed profile can still differ substantially in their local spatial gradients. The larger-$y$ regions generally display weaker gradients and a narrower range of reconstructed values.}

For the aorta, {the missing region at smaller $y$ than the known interval (to the left of the gray band in the profile plot)} intersects the highly turbulent flow generated by the systolic jet downstream of the aortic valve. The stochastic reconstructions consequently span a broad range and contain pronounced local extrema. Although averaging smooths these sharp variations, the ensemble mean recovers the high-velocity character of this region and its transition into the observed strip. {At larger $y$, to the right of the known interval,} where the GT profile is smoother and has a lower amplitude, the ensemble is narrower and its mean follows the gradual decay of the velocity toward the edge of the fluid domain.

The same dependence on the local flow structure is visible for the ventricle. {At smaller $y$ than the known interval,} the cut crosses the most intense and heterogeneous part of the field: the ensemble spread is comparatively broad, while its mean reproduces the rise and high-velocity plateau of the GT profile. {At larger $y$ than the known interval,} both the GT velocity and the sample-to-sample variability are smaller, and the ensemble mean follows the low-amplitude profile toward the boundary. The contrast between the two sides of each strip shows that the variability of the conditional reconstructions is governed by the local complexity of the missing flow, rather than solely by the distance from the available observation.

\subsection{GPR baseline} 
{As a probabilistic baseline, we use Gaussian process regression (GPR), which represents the observed and missing field values through a joint Gaussian distribution whose mean and correlations are estimated from the training data. Conditioning on the observed strip gives a mean reconstruction and a distribution from which alternative completions can be sampled. The formulation and implementation are detailed in Appendix~\ref{sec:gpr_method}.}

{Our implementation uses dense covariance matrices and Cholesky factorization to evaluate the Gaussian conditional distribution. Storing a covariance matrix for $D$ scalar field values requires $\mathcal{O}(D^2)$ memory, while factorizing the regularized covariance of the $d_m$ observed values requires $\mathcal{O}(d_m^3)$ operations~\cite{rasmussen2006gaussian}. To keep these matrix operations manageable in the present implementation, we restrict the direct comparison between DM and GPR to the same reduced aortic domain of $32\times256$ points, outlined in red in Fig.~\ref{fig:gpr_stripe}. For this comparison, a separate DM is trained on the same $10^4$ slices cropped to the reduced domain used by GPR, following the protocol in Appendix~\ref{sec:diffusion_method}. Both methods therefore use the same training data and observation configuration on the same domain.
Figure~\ref{fig:gpr_stripe}(a--d) shows that the DM distributions encompass the large GT excursions across much of the missing region. The GPR distributions are more concentrated and, at several positions, assign little visible density to the GT values.}

{Figure~\ref{fig:gpr_stripe}(a,c) compares the DM and GPR speed profiles along the same vertical cut through the reduced aortic domain; panels (b,d) show the corresponding absolute speed gradients. Both methods share the same known interval and GT profile.} {For the speed profiles, the mean curve is $\langle v\rangle$, evaluated over the stochastic reconstructions. The colored densities retain the pointwise variability lost on averaging.}

{In the reconstructed region at smaller $y$ than the known interval (to the left of the gray band in the profile plots),} the GT exhibits a high and strongly fluctuating velocity magnitude. The DM mean captures its characteristic level and follows its principal variations, despite smoothing some of the sharpest extrema. In contrast, the GPR mean is biased toward substantially lower values over most of this region and does not reproduce the large GT fluctuations. {In the reconstructed region at larger $y$ (to the right of the gray band),} where the GT velocity is lower, the DM mean again remains closer to the reference profile, whereas GPR systematically overestimates it. Thus, although the DM realizations span a broad conditional range, their mean is appreciably less biased than the GPR mean on both sides of the observed strip.

{The gradient distributions in Fig.~\ref{fig:gpr_stripe}(b,d) provide a complementary view of reconstruction variability. The DM spans a broader range of absolute gradients in the missing regions, while the GPR distribution is more concentrated and varies markedly with position. The differences between the black GT curve and the cyan mean curves also show that matching the broad speed profile does not guarantee recovery of the local gradients of an individual reference field.}

{Figure~\ref{fig:gpr_stripe}(e--g) extends the comparison beyond repeated reconstructions of the single reference slice shown in (a--d), using an ensemble of different GT images and their corresponding reconstructions. Panel (e) compares the speed PDFs normalized by the standard deviation of each ensemble. The DM more closely follows the GT distribution over much of the plotted range, including its high-speed shoulder, whereas the GPR distribution shows a broader discrepancy at intermediate speeds and in the tail. Panel (f) compares the normalized PDFs of the in-plane gradient magnitude $|\nabla v|$. Here, too, the DM follows the GT more closely over much of the range; the inset highlights differences in the peak region, while the full plot exposes differences in the tails.}

{Panel (g) reports the flatness of vertical speed increments using the same definition as in Fig.~\ref{fig:unconditional_snapshots}{(h)}. Flatness is calculated separately for each image. Both methods reproduce its overall decrease with separation. At small separations, the DM curve lies closer to the GT than the GPR curve, while at larger separations all three curves approach one another.}

{Figure~\ref{fig:gpr_stripe}(h) complements this ensemble comparison with pointwise error PDFs for the fixed observation used in the cut plots. The two positions, $y=0.005\,\mathrm{m}$ and $y=0.025\,\mathrm{m}$, lie outside the known strip, farther from and closer to its boundary, respectively. For each reconstruction, we evaluate the absolute difference between reconstructed and GT speeds and divide it by a single GT RMS speed computed over all grid points of the $32\times256$ crop, including the known region. At both locations, the DM achieves a lower mean absolute speed error than GPR, with a more pronounced improvement farther from the known strip. Its error distributions extend closer to zero but have broader tails, whereas the narrower GPR distributions remain concentrated at nonzero errors. Thus, a smaller spread does not necessarily imply more accurate reconstruction.}

{For the smoother ventricular flow, GPR performs more similarly to the DM on the reduced domain (data not shown).}

\section{Conclusions, limitations and future perspectives}
\label{sec:conclusions}

{This study demonstrates two complementary capabilities of diffusion models for patient-specific cardiovascular flow: generating diverse complete fields and reconstructing large missing regions from limited observations.} Separate models were trained for the aorta and ventricle, allowing the markedly different geometries and flow regimes of the two regions to be represented. In the unconditional setting, the generated aortic slices reproduce the variations in geometry and velocity associated with the cardiac cycle, as well as the one-point distribution and scale-dependent intermittency of the velocity magnitude. The comparison of velocity-gradient fields further shows that the generated ensemble retains a degree of structural diversity comparable to that of {distinct training fields}.

{For full-slice conditional reconstruction, no retraining is needed: the same models reconstruct }{test fields} { from a central observation containing only about $30\%$ of the complete slice.} Despite the removal of two large regions, the conditional samples recover the overall fluid-domain geometry and the dominant organization of both aortic and ventricular flows. The remaining variability is concentrated in the least constrained and most complex regions, particularly around the aortic valve and the systolic jet, where several small-scale configurations can be compatible with the same observation. {Along the full-slice profiles, the DM ensemble mean retains the principal features of the reference flow while averaging over local fluctuations. The direct comparison with GPR is restricted to the separately trained models on the reduced aortic domain. There, the DM ensembles encompass a broader range of the GT speed and gradient excursions, whereas GPR realizations can concentrate away from the reference signal.} {The ensemble PDFs and flatness further test the recovery of flow variability across different reference fields, with the DM generally closer to the GT distribution shapes and small-separation flatness in the reported comparisons.} {The study demonstrates DM-based reconstruction on complete aortic and ventricular slices and compares reconstruction quality with GPR on a common reduced domain. In that comparison, the DM represents a broader range of plausible completions and better encompasses the GT excursions.}

The present work is a controlled proof of concept rather than a direct reconstruction of clinical measurements. The database represents one healthy anatomy and {a heart rate of $60\,\mathrm{bpm}$, chosen as a healthy reference condition}; the model operates on two-dimensional slices, and the available region is prescribed exactly rather than being affected by measurement noise and {acquisition} artifacts. Extending the training data to different anatomies, physiological conditions, and pathological flow regimes will therefore be necessary to assess generalization and clinical relevance.

{The present results do not establish whether the architecture and training strategy used here have sufficient capacity to represent full three-dimensional cardiovascular fields or their joint evolution in time. Although physics-constrained diffusion models have demonstrated single-time generation of three-dimensional turbulent fields~\cite{li2026_physically_consistent_3d}, extending this capability to moving cardiac geometries and complete heartbeat sequences remains to be assessed. The required number of training configurations, their temporal coverage, and the associated simulation, storage, and model-training costs are also unknown; generating sufficiently diverse, fully resolved cardiovascular data may become a limiting factor despite GPU acceleration~\cite{viola2022fsei_gpu,viola2023high}. For the present simulations at $60\,\mathrm{bpm}$, each cardiac cycle requires approximately $30\,\mathrm{h}$ of wall-clock time on two NVIDIA A100 GPUs. Furthermore, our generative model does not explicitly enforce incompressibility or the no-slip condition on the moving walls and valve leaflets. These properties are enforced in the simulation database, but learning from such data does not guarantee their exact satisfaction in generated fields. Incorporating physical constraints into diffusion sampling is a promising direction~\cite{li2026_physically_consistent_3d}, although enforcing velocity compatibility with deforming boundaries would require information about their motion and additional validation. Finally, the spread of the conditional ensemble represents variability under the learned model; it has not been calibrated as uncertainty for clinical measurements.}

{The next step is to apply this framework to clinical flow measurements from different imaging modalities (e.g., color-Doppler echocardiography and flow MRI). Conditioning would then need to account for what each instrument actually measures, its spatial and temporal resolution, and regions affected by noise or missing data. Numerical simulations could support this transition by generating paired synthetic measurements and complete velocity fields under varied acquisition settings and artifact levels. Such data could be used to pretrain models before adaptation to clinical measurements.}

A further development is to replace independent slice reconstruction with a spatiotemporal model of the complete heartbeat. {Conditioning jointly on neighboring frames, cardiac anatomy, and the phase of the cycle could improve temporal consistency and exploit the organization of cardiovascular flow, but its feasibility and statistical accuracy require dedicated three-dimensional and time-resolved tests.} Beyond improving reconstruction, the generative ensemble could then be used to estimate clinically relevant quantities together with their uncertainty, such as jet extent, regurgitant volume, pressure gradients, or vortex characteristics.

\begin{acknowledgments}
This work was supported by the Italian Ministry of University and Research (MUR) under the FARE program (No. R2045J8XAW), project “Smart-HEART”, and by the European Research Council (ERC) under the European Union’s Horizon 2020 research and innovation program (Grant Agreement No. 882340).
Luca Biferale and Fabio Guglietta acknowledge funding from Tor Vergata University project AI4HEART. 
Francesco Viola acknowledges financial support by the European Research Council (ERC) under the European Union’s Horizon Europe research and innovation program, Project CARDIOTRIALS (Grant Agreement No. 101039657).
\end{acknowledgments}
\section*{Data availability}
The datasets and analysis code supporting the findings
of this study are available from the corresponding
author upon reasonable request.
\appendix

\begin{figure}[t!]
    \centering
    \includegraphics[width=1.\linewidth]{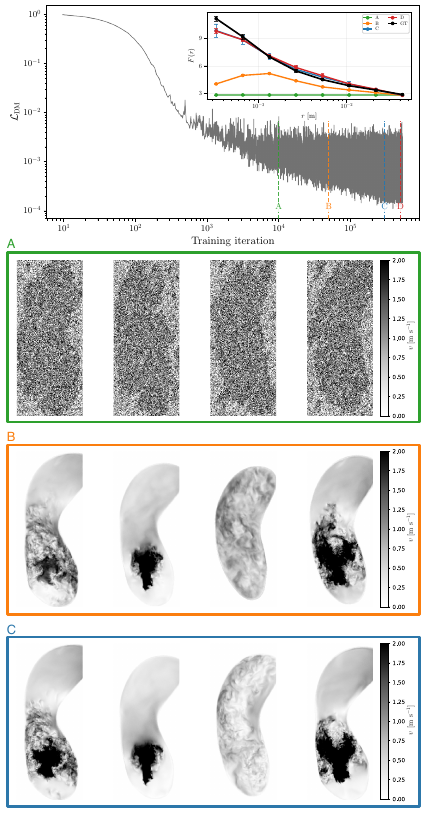}
    \caption{{Evolution of unconditional aortic generation during training. Top: noise-prediction loss $\mathcal{L}_{\mathrm{DM}}$ versus optimization iteration, on logarithmic axes. Colored vertical lines mark checkpoints A (10K, green), B (50K, orange), C (300K, blue), and D (500K, red). The inset compares the flatness $F(r)$ of vertical speed increments generated at these checkpoints with GT (black), using the definition and split-sample summary described for Fig.~\ref{fig:unconditional_snapshots}. The three rows below show four representative generated fields at checkpoints A, B, and C, respectively. Grayscale represents the velocity magnitude $v$ in $\mathrm{m\,s^{-1}}$, using the same color scale in all rows.}}
    \label{fig:training}
\end{figure}

\section{{Diffusion model (DM)}}
\label{sec:diffusion_method}

We use a DM~\cite{ho2020denoising} to learn the distribution of the complete cardiovascular slices. {The unconditional generative backbone follows the approach developed for turbulent signals by Li and coworkers~\cite{li2024synthetic}.} Separate models are trained for the aorta and the ventricle: they use identical architecture and training hyperparameters, but have independently optimized weights. {For each full-slice anatomical domain, the same trained model is used for unconditional generation and conditional reconstruction, without further optimization for conditioning. The reduced-domain DM used in the GPR comparison is trained separately on cropped slices, as described in Sec.~\ref{sec:results}; the no-retraining statement refers to conditioning a model within its own trained domain.}

Let $\mathbf{x}_0$ denote a sample, normalized to $[-1,1]$ {as in Sec.~\ref{sec:fsei_database}}, drawn from the training distribution. The forward diffusion process constructs a sequence $\mathbf{x}_1,\ldots,\mathbf{x}_N$ by progressively adding Gaussian noise according to
\begin{equation}
q(\mathbf{x}_n|\mathbf{x}_{n-1})=
\mathcal{N}\!\left(\sqrt{1-\beta_n}\,\mathbf{x}_{n-1},\beta_n\mathbf{I}\right),
\end{equation}
where $\beta_n$ sets the noise variance at step $n$. The implementation uses $N=2000$ diffusion steps and a linear noise schedule; in the notation above, $n$ labels the corresponding discretized noise level. Defining $\alpha_n=1-\beta_n$ and $\bar{\alpha}_n=\prod_{m=1}^{n}\alpha_m$, an arbitrary noisy state can be sampled directly as
\begin{equation}
\mathbf{x}_n=\sqrt{\bar{\alpha}_n}\,\mathbf{x}_0+
\sqrt{1-\bar{\alpha}_n}\,\boldsymbol{\epsilon},
\qquad \boldsymbol{\epsilon}\sim\mathcal{N}(\mathbf{0},\mathbf{I}).
\label{eq:forward_diffusion}
\end{equation}
For sufficiently large $N$, $\mathbf{x}_N$ approaches an uncorrelated Gaussian field and contains negligible information about the original sample.

The reverse process is parameterized by a neural network $\boldsymbol{\epsilon}_{\theta}(\mathbf{x}_n,n)$ that estimates the noise contained in $\mathbf{x}_n$. The aortic and ventricular models use the same network architecture and training hyperparameters, with four input channels (i.e., $u_x$, $u_y$, $u_z$, and $\chi$), of size $176\times416$ and $304\times272$, respectively. The denoiser is implemented as a multiscale U-Net with 64 base channels, three residual blocks at each resolution level, and channel multipliers $(1,2,3,4)$, corresponding to feature widths $(64,128,192,256)$. {The U-Net includes an attention layer in its bottleneck.} Its parameters $\theta$ are determined by minimizing the standard noise-prediction objective
\begin{equation}
\mathcal{L}_{\mathrm{DM}}=
\mathbb{E}_{{\mathbf{x}_0,n,\boldsymbol{\epsilon}}}
\!\left[
\left\|
\boldsymbol{\epsilon}-
{\boldsymbol{\epsilon}_{\theta}(\mathbf{x}_n,n)}
\right\|_2^2
\right].
\label{eq:diffusion_loss}
\end{equation}
Each anatomical model is trained independently for $3\times10^5$ iterations with a learning rate of $10^{-4}$ and a batch size of 16. {The choice of $3\times10^5$ optimization iterations is supported by the aortic training-checkpoint comparison in Fig.~\ref{fig:training}, where extending training to $5\times10^5$ iterations produces little further change in the generated flatness.}

For unconditional generation, sampling starts from $\mathbf{x}_N\sim\mathcal{N}(\mathbf{0},\mathbf{I})$ and repeatedly applies the learned reverse transitions until a complete field $\mathbf{x}_0$ is obtained. {For the training-checkpoint comparison in Fig.~\ref{fig:training}, $5\times10^3$ unconditional samples are generated independently at each of the four displayed checkpoints. Flatness is computed separately for each retained sample and summarized using the same split-sample procedure as in Fig.~\ref{fig:unconditional_snapshots}(h).} For conditional reconstruction, let $\mathbf{M}$ be the binary observation mask, equal to one in the known region and zero elsewhere {(the known region is shown in green in Fig.~\ref{fig:sketch_recon_aorta})}. This mask is distinct from the fluid-domain indicator $\chi$, which remains one of the four physical channels. {Conditioning is performed using diffusion posterior sampling (DPS)~\cite{chung2023diffusion}. Let $\widehat{\mathbf{x}}_0(\mathbf{x}_n,n)$ denote the denoised estimate of the complete field inferred from the current noisy state, and let $\mathbf{x}^{\mathrm{obs}}$ contain the supplied values in the known strip. In the implementation used here, the discrepancy and guided update are}
\begin{equation}
{\begin{aligned}
R_n &= \frac{\left\|\mathbf{M}\odot\bigl[\widehat{\mathbf{x}}_0(\mathbf{x}_n,n)
-\mathbf{x}^{\mathrm{obs}}\bigr]\right\|_2}{\sqrt{N_{\mathrm{obs}}}},\\
\mathbf{x}_{n-1} &= \widetilde{\mathbf{x}}_{n-1}
-\lambda_{\mathrm{DPS}}\nabla_{\mathbf{x}_n} R_n,
\end{aligned}}
\label{eq:diffusion_conditioning}
\end{equation}
{where $\widetilde{\mathbf{x}}_{n-1}$ is the proposal from the reverse-diffusion step, $\odot$ denotes pointwise multiplication, and $N_{\mathrm{obs}}$ is the number of observed scalar entries, including all four channels. The discrepancy is evaluated in the normalized variables used by the model, and the gradient is propagated through the denoised estimate. The guidance strength is $\lambda_{\mathrm{DPS}}=128$. This correction encourages agreement with the observation during sampling. The same spatial mask is applied to all four channels, so that both the unobserved velocity components and the corresponding fluid-domain geometry are reconstructed.}

Because the reverse dynamics is stochastic, independent initial noise fields produce different completions for the same observation. {Their mean provides an average reconstruction, while their spread describes the variability under the learned conditional model.} {For the full-slice conditional results presented in Sec.~\ref{sec:results},} one aortic and one ventricular slice {from their respective test datasets} are each reconstructed $K=4\times10^3$ times using independent initial noise fields. 

{Figure~\ref{fig:training} motivates the choice of $3\times10^5$ training iterations for the aortic model. The loss decreases rapidly during the initial stage and continues to decline more slowly thereafter. However, a low noise-prediction loss alone does not establish that generated fields reproduce the flow statistics. At 10K iterations, the samples remain noise-like, whereas by 50K the aortic geometry and organized flow structures emerge but the flatness remains substantially below the GT at small separations. At 300K, the generated fields exhibit finer spatial structure and the flatness is much closer to the GT curve. Extending training to 500K produces little further change in $F(r)$ over the resolved separations. We therefore adopt the 300K checkpoint as a compromise between training cost and the recovery of the spatial statistics examined here.}

\section{{Gaussian process regression baseline}}
\label{sec:gpr_method}

Gaussian process regression (GPR) is used as a probabilistic baseline, following the construction adopted for gappy turbulent signals in Ref.~\cite{li2025stochastic}. In the present spatial formulation, the three velocity components {and the fluid-domain indicator $\chi$} of a two-dimensional slice are concatenated into a single vector and partitioned into the measured values $\mathbf{x}_{m}$, located within the known strip, and the missing values $\mathbf{x}_{g}$. These two sets of values are modeled as a jointly Gaussian random vector,
\begin{equation}
\begin{bmatrix}
\mathbf{x}_{m}\\
\mathbf{x}_{g}
\end{bmatrix}
\sim
\mathcal{N}\!\left(
\begin{bmatrix}
\boldsymbol{\mu}_{m}\\
\boldsymbol{\mu}_{g}
\end{bmatrix},
\begin{bmatrix}
\mathbf{C}_{mm} & \mathbf{C}_{mg}\\
\mathbf{C}_{gm} & \mathbf{C}_{gg}
\end{bmatrix}
\right).
\label{eq:gpr_joint}
\end{equation}
Here, $\boldsymbol{\mu}_{m}$ and $\boldsymbol{\mu}_{g}$ are the {mean field-value vectors, including the velocity components and the domain-indicator channel, restricted to the measured and missing regions}, while the four blocks of $\mathbf{C}$ contain both spatial and cross-component covariances. The means and covariance entries are estimated from the same $10^4$ training slices used for the DM.

Conditioning Eq.~\eqref{eq:gpr_joint} on the available values $\mathbf{x}_{m}$ gives
\begin{align}
p_{\mathrm{GPR}}(\mathbf{x}_{g}|\mathbf{x}_{m})
&=\mathcal{N}(\boldsymbol{\mu}_{g|m},\mathbf{C}_{g|m}), \nonumber\\
\boldsymbol{\mu}_{g|m}
&=\boldsymbol{\mu}_{g}
+\mathbf{C}_{gm}(\mathbf{C}_{mm}+\lambda_{\mathrm{jit}}\mathbf{I})^{-1}
(\mathbf{x}_{m}-\boldsymbol{\mu}_{m}), \nonumber\\
\mathbf{C}_{g|m}
&=\mathbf{C}_{gg}
-\mathbf{C}_{gm}(\mathbf{C}_{mm}+\lambda_{\mathrm{jit}}\mathbf{I})^{-1}\mathbf{C}_{mg}.
\label{eq:gpr_conditional}
\end{align}
Here, a positive jitter coefficient $\lambda_{\mathrm{jit}}$ is added to the diagonal of $\mathbf{C}_{mm}$ to improve its numerical conditioning and stabilize the Cholesky factorization~\cite{rasmussen2006gaussian}. Its value is set relative to the mean marginal variance of the measured vector,
\begin{equation}
\lambda_{\mathrm{jit}}
=10^{-2}\,\frac{\operatorname{tr}(\mathbf{C}_{mm})}{d_m},
\label{eq:gpr_jitter}
\end{equation}
where $d_m$ is the dimension of $\mathbf{x}_m$. The conditional mean $\boldsymbol{\mu}_{g|m}$ is the GPR mean reconstruction, and the conditional covariance $\mathbf{C}_{g|m}$ {specifies the variability of the Gaussian conditional model and is used to draw stochastic reconstructions.} The three velocity components are reconstructed jointly {with $\chi$}, thereby retaining their cross-correlations, and are conditioned using the same spatial observation mask as the DM; the measured velocity values are retained exactly. Velocity magnitudes are subsequently computed from the three reconstructed components. For each selected test slice, $4\times10^3$ GPR realizations are drawn so that the GPR and DM statistics are evaluated from ensembles of the same size. {For the ensemble statistics of the reduced aortic domain in Fig.~\ref{fig:gpr_stripe}(e--g), we additionally reconstruct different reference images with both methods and compare their distributions with those of the corresponding GT fields.}

\bibliography{main}

\end{document}